\documentclass{cjpsuppl}
\usepackage{epsfig}
\begin{document}
\title{Jet quenching}
\authori{N\'estor Armesto}
\addressi{Department of Physics, CERN, Theory Division,
CH-1211 Gen\`eve 23, Switzerland}
\authorii{}    \addressii{}
\authoriii{}   \addressiii{}
\authoriv{}    \addressiv{}
\authorv{}     \addressv{}
\authorvi{}    \addressvi{}
\headtitle{Jet quenching}
\headauthor{N\'estor Armesto}
\lastevenhead{N\'estor Armesto: Jet quenching}
\refnum{}
\daterec{;\\final version }
\suppl{A}  \year{2004} \setcounter{page}{1}
\maketitle

\begin{abstract}
Jet quenching studies play a prominent role in our current understanding of
ultra-relativistic
heavy ion collisions. In this review
I first present the available formalism to compute
medium-induced gluon radiation. Then I discuss its effect on single particle
spectra, with dedicated attention to the case of the radiating parton being a
massive quark. Next I examine more differential observables like jet shapes
and multiplicities, and the consequences of flow on radiative
energy loss. I conclude
with
some remarks.
\end{abstract}

\section{Introduction and formalism}
\label{sec1}

\begin{figure}[htb]
\begin{center}
\epsfig{file=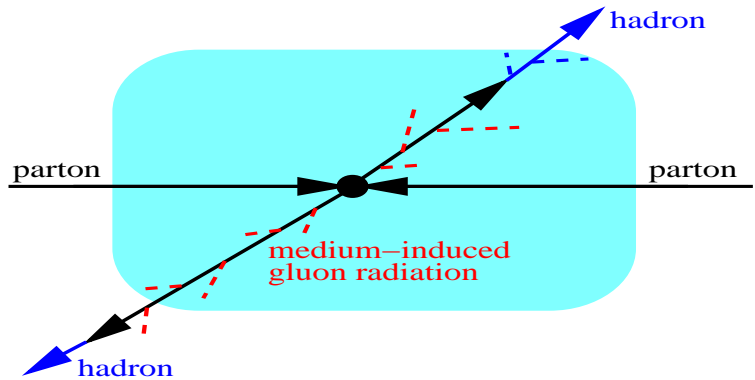,width=10cm,height=5cm}
\end{center}
\vskip -0.8cm
\caption{Jet quenching by radiation off fast partons produced inside a medium.}
\label{fig1}
\end{figure}

Jet quenching (Fig. \ref{fig1}) plays a central role in heavy ion studies
\cite{muller,harris}. It constitutes the standard
explanation of the large transverse momentum suppression of
single particle spectra in nucleus-nucleus collisions with respect to the
expectation from nucleon-nucleon and the disappearance of back-to-back
correlations, observed at RHIC (see the reviews
\cite{Baier:2000mf,Kovner:2003zj,Gyulassy:2003mc,Salgado:2003qc,Vitev:2004bh}
for the existing formalism
and comparisons with data). This phenomenon is taken as a prominent indication
of the creation of a dense partonic system in ultra-relativistic
heavy ion collisions
\cite{muller,harris}.

Historically, the first proposal \cite{Bjorken:1982tu}
was based on elastic scattering. Later on it was recognized that
bremsstrahlung was the dominant process for high energies of the parton
traversing the medium, first in a
QED form \cite{Gyulassy:1993hr} and then taking into account the rescattering
of the emitted gluon \cite{Baier:1994bd}.
It is now widely believed that for energies of the parton
(equivalent to transverse momentum
in the central rapidity region) greater than $5\div 10$ GeV,
hadronization takes place outside any medium produced in the collision
\cite{Wiedemann:2004wp} and
thus gluon radiation dominates the energy loss. Medium-induced gluon radiation
implies an energy degradation of the leading parton, a broadening of the
associated parton shower and an increase of the associated hadron multiplicity.

\begin{figure}[htb]
\begin{center}
\epsfig{file=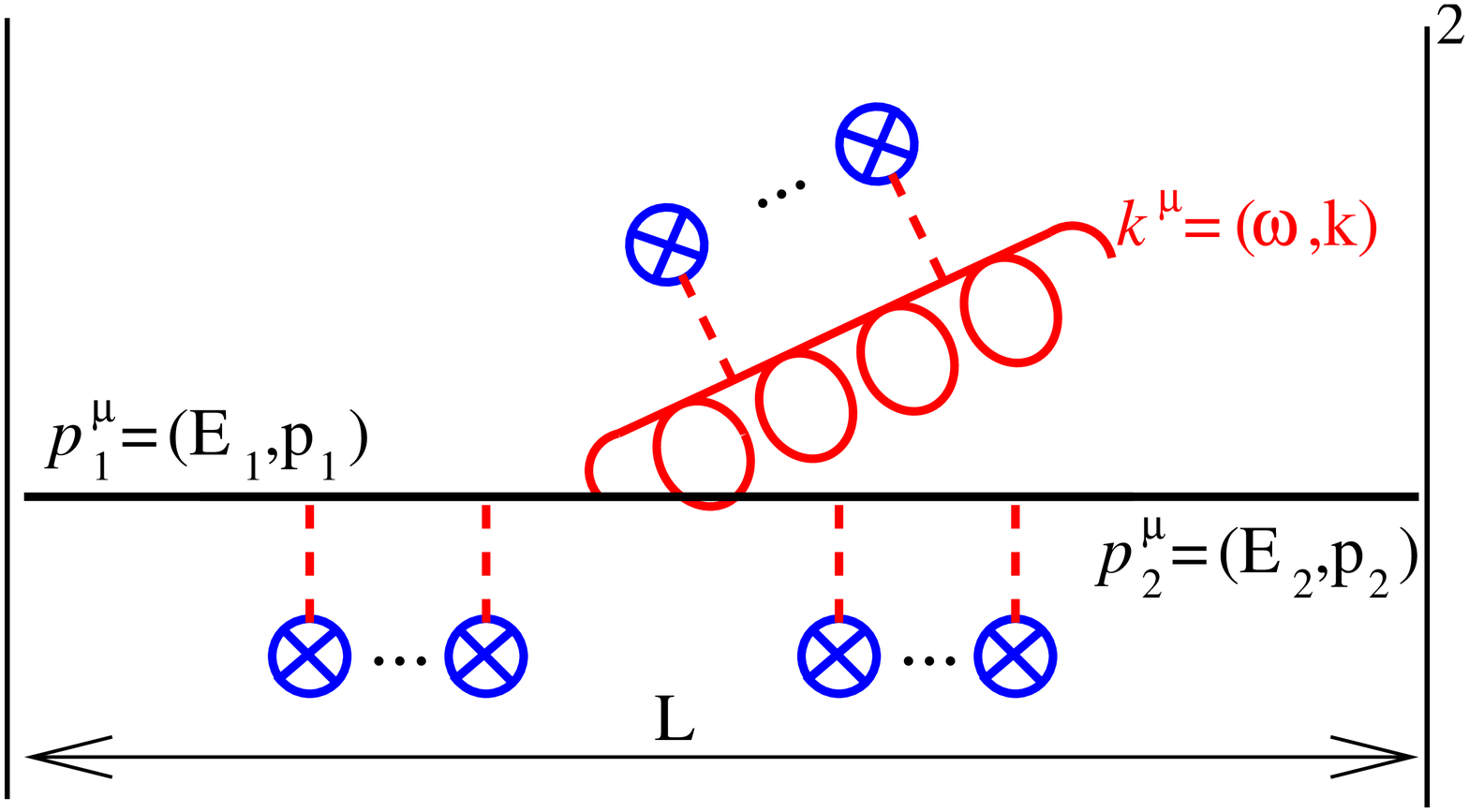,width=12cm,height=5cm}
\vskip 0.1cm
\epsfig{file=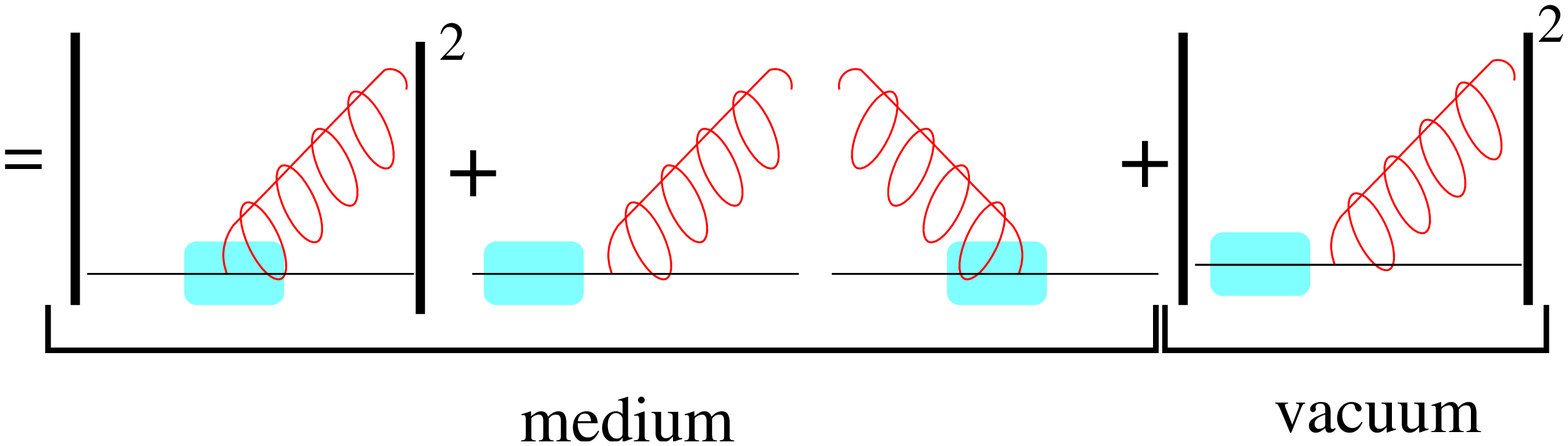,width=12cm,height=4cm}
\end{center}
\vskip -0.8cm
\caption{Diagrams relevant to medium-induced energy loss of a fast parton in a
medium, with the kinematical variables. Three types of probabilities
appear: the vacuum term in which the gluon is both emitted in the amplitude and
absorbed in the complex conjugate amplitude outside the medium, the one in
which both absorption and emission take place in the medium, and the
crossed term.}
\label{fig2}
\end{figure}

The available formalism (its most general formulation can be found in
\cite{Zakharov:1996fv,Wiedemann:2000za})
takes into account the rescattering of the
incoming and outgoing radiating parton and of the radiated gluon with the
medium, see Fig. \ref{fig2}.
Interference and mass effects on the energy spectrum of radiated gluons
$\omega\frac{dI_{\rm
medium}}{d\omega\, d{\bf k}_\perp}$ are given by the crossed term in Fig.
\ref{fig2} and contained in an exponential $\exp{\left(-\Delta
z\,{k_\perp^2+x^2m^2\over 2\omega}\right)}$,
with $m$ the mass of the radiating parton and $x={\omega \over E}\ll 1$.
The information about the medium comes in the product
density of the medium times cross section for the rescattering of the gluon
inside the medium. Different approximations are taken for this product, the
most popular one being $n(z)\sigma(r)\propto \hat q(z) r^2$.
The energy loss turns out to be energy-independent and, due to the fact that
the dominant process is the rescattering of the radiated gluon, proportional
to the length of the medium squared, $\Delta E \simeq
\int d\omega\,
\omega {dI \over d\omega} \propto \alpha_s
C_R \omega_c = \alpha_s
C_R \hat q L^2/2
$, $\omega_c=\hat q L^2/2$ and $\hat q=\mu^2/\lambda$ the transport coefficient
defined as the mean transverse momentum squared transferred from the medium
to the gluon
per unit mean free path
(see \cite{Baier:2002tc} for simple arguments).

\section{Mean energy loss and single particle spectra}
\label{sec2}

Two ways, at the level of either the $p_\perp$-spectrum or the fragmentation
functions,
have been proposed to compute the
medium-modified particle spectrum
\cite{Guo:2000nz,Baier:2001yt,Wang:2002ri}:
\begin{equation}
{d\sigma^{\rm medium}(p_\perp)\over dp_\perp^2}=\int d\epsilon \, P(\Delta E)
\,{d\sigma^{\rm vacuum}(p_\perp+\Delta E)\over dp_\perp^2}\ ;
\label{eq1}
\end{equation}
\begin{equation}
D_{h/p}^{\rm medium}(z,Q^2)=\int d\epsilon \, {P(\Delta E) \over 1-\epsilon}\,
D_{h/p}^{\rm vacuum}\left({z \over 1-\epsilon},Q^2\right),
\ \epsilon={\Delta E \over E{\small \simeq p_\perp (y=0)}}.
\label{eq2}
\end{equation}
From these expressions, a strong influence of the form of the partonic
spectrum is evident. The function $P(\Delta E)$ gives the probability for the
fragmenting parton to loose some energy $\Delta E$ and is called the quenching
weights. The usual approximation to compute these weights is the Poissonian one
which considers each subsequent
gluon emission from the parton as independent
\cite{Salgado:2003gb}. The centrality dependence of the suppression of the
transverse momentum spectra can be understood in terms of nuclear
geometry \cite{andrea,Drees:2003zh}. The medium produced in the collision
expands and its dynamical dilution
can be absorbed in a redefinition of
$\hat q$ \cite{Baier:1998yf,Salgado:2002cd,Gyulassy:2000gk}:
\begin{equation}
\hat q_{eff}(L)={2\over L^2} \int_{\tau_0}^L d\tau \, (\tau-\tau_0) \hat
q(\tau),
\label{eq3}
\end{equation}
with $\hat
q(\tau)$ the transport coefficient depending on the proper time.
With this formalism, a good agreement with the existing experimental data is
found, see e.g. \cite{Dainese:2004te,Eskola:2004cr} for recent comparisons.
Several conclusions have been extracted: First, the medium is opaque at RHIC,
in the sense that there is quite a large insensitivity to the value of $\hat
q$, and thus to the density, provided it is high enough. The emission takes
place from the surface of the medium which naturally explains the absence of
back-to-back correlations and the approximate scaling with number of
participants. Second, uncertainties at small $p_\perp$ due to finite energy
constraints, imposed a posteriori on the theoretical calculations which are
done in the high-energy limit, are clearly
visible. As a final remark, predictions for quenching at different energies
are
usually done 
by rescaling
$\hat q$ according to multiplicities
\cite{Vitev:2004gn,Adil:2004cn,Wang:2004yv}.

\section{Gluon radiation off massive quarks}
\label{sec3}

\begin{figure}[htb]
\begin{center}
\epsfig{file=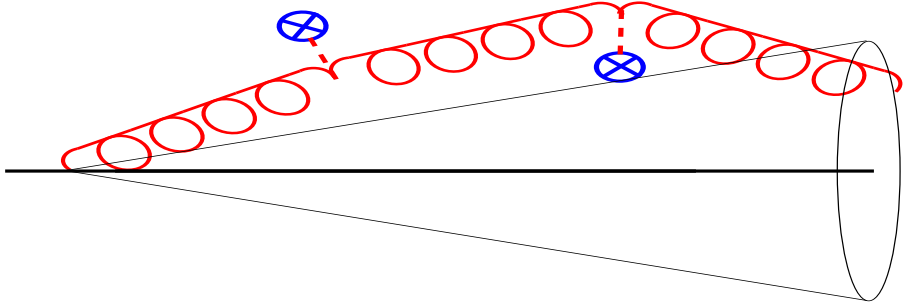,width=12cm,height=4cm}
\end{center}
\vskip -0.8cm
\caption{Movement of the emitted gluon into the dead
cone due to its multiple scattering.}
\label{fig3}
\end{figure}

Gluon radiation in the vacuum is modified by
a mass of the parent quark. Radiation for angles $\theta<m/E$ is
suppressed, which constitutes the so-called
dead cone effect \cite{Dokshitzer:1991fc}. It leads to the modification of the
spectrum of the radiated gluon
\begin{equation}
{1\over
{\bf k}_\perp^2} \longrightarrow {1\over
{\bf k}_\perp^2} \left[{{\bf k}_\perp^2 \over
{\bf k}_\perp^2+\left({m\omega\over E}\right)^2}\right]^2
\equiv {1\over
{\bf k}_\perp^2}\ F\left({\bf k}^2_\perp,{m\omega \over E}\right)\ .
\label{eq4}
\end{equation}
Dokshitzer and Kharzeev \cite{Dokshitzer:2001zm}
proposed that
medium-induced gluon radiation
is reduced by the same effect. In this first exploratory study, in absence of
a fully differential spectrum they considered
\begin{equation}
\omega\frac{dI^{m>0}_{\rm medium}}{d\omega} =
\omega\frac{dI^{m=0}_{\rm medium}}{d\omega}\ F\left(\langle{\bf
k}^2_\perp\rangle,{m\omega \over E}\right),\ \ \langle{\bf
k}^2_\perp\rangle \simeq \sqrt{\hat q \omega}.
\label{eq5}
\end{equation}
Naively one would expect that the gluon may move into the dead cone due to
multiple scattering, and thus the dead cone may be filled, see Fig.
\ref{fig3}. 
Technically, there is a competition between interference and rescattering
which demands numerical studies, done in
\cite{Djordjevic:2003qk,Djordjevic:2003zk,Zhang:2003wk} for fragmentation
functions and the mean energy loss, and in \cite{Armesto:2003jh} for the
fully
differential spectrum.

\begin{figure}[htb]
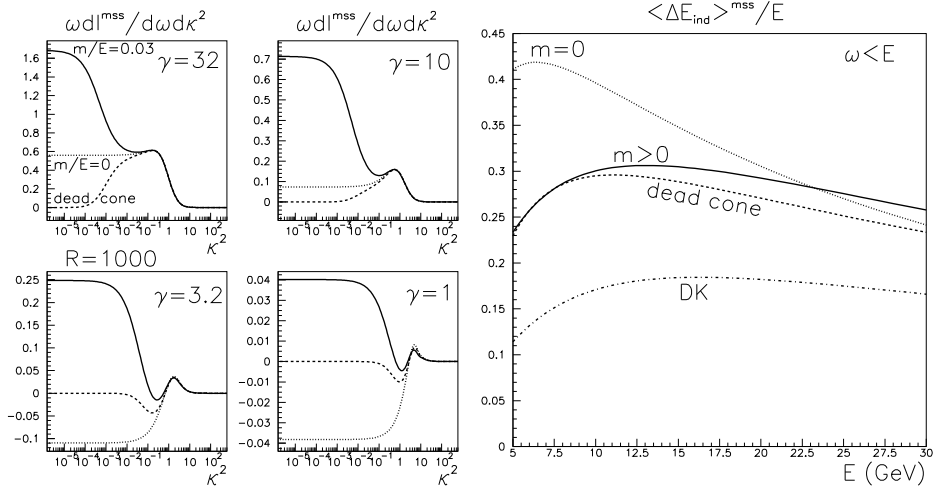

\begin{center}
\epsfig{file=fig6.epsi,width=6cm}\hskip 0.3cm\epsfig{file=fig9right.epsi,width=6cm}
\end{center}
\vskip -0.8cm
\caption{Left: differential spectrum of radiated gluons versus $\kappa^2={\bf
k}_\perp^2/(\hat q L)$ for $R=\omega_cL=1000$, $\omega_c=\hat q L^2/2$, for
different $\gamma=\omega_c/\omega$. Right: relative mean induced
energy loss $\langle \Delta
E_{\rm ind}\rangle/E$
for $L=6$ fm and $\hat{q}=
0.8$ GeV$^2$/fm, versus $E$.}
\label{fig4}
\end{figure}

In Fig. \ref{fig4} it is shown \cite{Armesto:2003jh} on the left the
differential spectrum of radiated gluons, and on the right the mean induced
energy loss
$\langle \Delta
E_{\rm ind}\rangle
  = \int_0^E d\omega\,   \omega\frac{dI_{\rm medium}}{d\omega}=\Delta E$
for parameters extracted from a comparison with RHIC data
\cite{Salgado:2003gb}. Some conclusions can be drawn:
The dead cone is filled, but it constitutes a small fraction of the
available phase space. The energy loss for charm quarks at RHIC is
a factor $\sim 2$
smaller
than for light quarks, but should still be observable.  And the uncertainties
due to energy constraints imposed a posteriori are significant. The
experimental situation on charm in AuAu collisions
at RHIC \cite{Adcox:2002cg,Adler:2004} is
unclear, as the available single electron spectra constrain weakly the energy
loss \cite{Batsouli:2002qf} and the measured $p_\perp$ is small so
hadronization effects may play a significant role.

\section{More differential observables}
\label{sec4}

\begin{figure}[htb]
\begin{center}
\epsfig{file=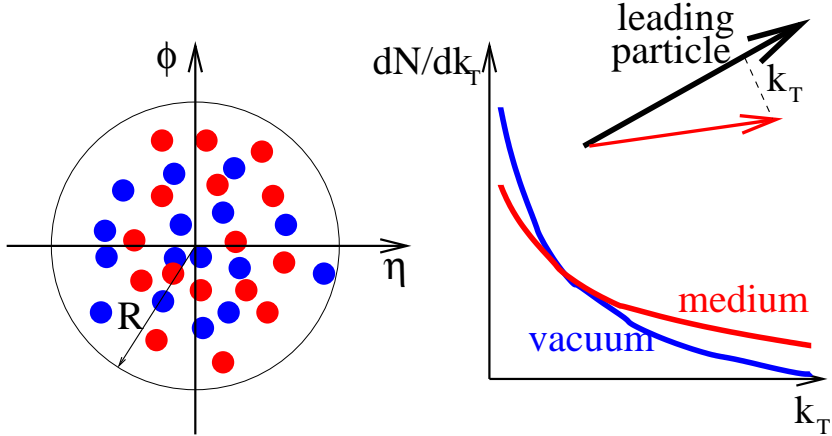,width=11cm}
\end{center}
\vskip -0.8cm
\caption{Schematic effects of medium-induced gluon radiation on jet shapes.}
\label{fig6}
\end{figure}

Medium-induced radiation may modify the energy deposition (i.e. the
jet definition
and
profile) and the distribution of sub-leading particles from that in usual
fragmentation in the vacuum, see Fig. \ref{fig6}.
Such studies have been
performed in \cite{Salgado:2003rv}. The conclusions are: The jet
definition is stable, as most energy is deposited at small
$R=\sqrt{\eta^2+\phi^2}$ with $\eta$ and $\phi$ the pseudorapidity and
azimuth with respect to the center of the jet, which gives good chances to
measure jets in a heavy ion environment at the LHC. There is little
sensibility to the infra-red contribution from the
medium. And the background is
apparently under control (the vacuum contribution is taken from
\cite{Abbott:1997fc} and may be fixed at the LHC from pp and pA). These
studies may also have important consequences for RHIC, see below.

\begin{figure}[htb]
\vskip -0.2cm
\begin{center}
\epsfig{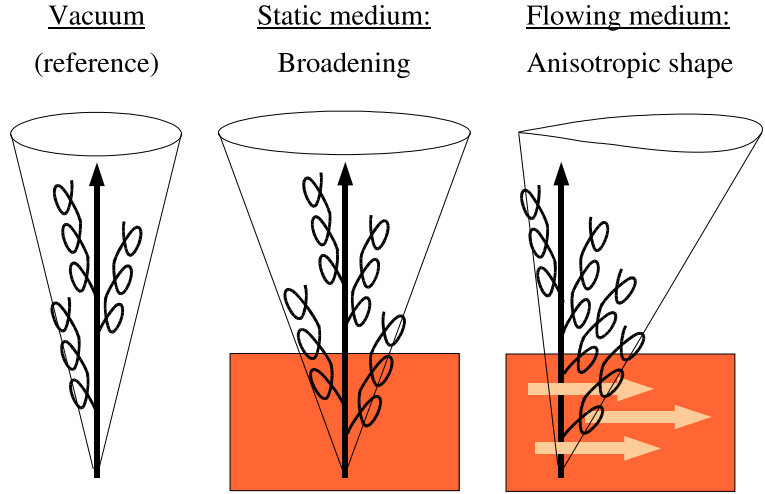}
\end{center}
\vskip -0.6cm
\caption{Gluon radiation in the vacuum, in a static medium and in a flowing
medium.}
\label{fig7}
\end{figure}

\begin{figure}[htb]
\begin{center}
\epsfig{file=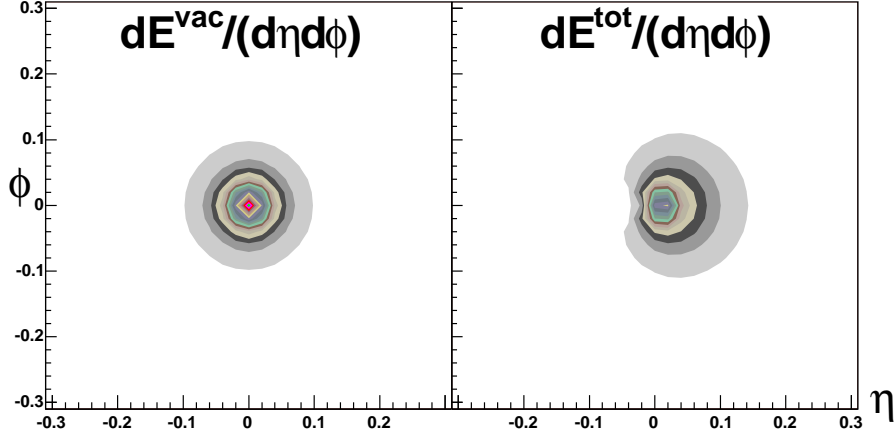,width=12cm}
\end{center}
\vskip -1.2cm
\caption{Energy deposition in the $\eta\times \phi$-plane for a 100 GeV gluon
jet with $\mu=1$ GeV, a directed momentum
component $q_0=\mu$ in the positive $\eta$-direction, and parameters
$n_0\, L\, \alpha_s
C_R = 1$ and $L=6$ fm (see \cite{Armesto:2004pt} for details), both
for the vacuum
(i.e. no medium,
plot on the left) and when medium is present (i.e. vacuum+medium,
plot on the right).}
\label{fig8}
\end{figure}

Until now, I have considered a medium which expands and dilutes but shows no
momentum anisotropy.
In heavy ion collisions, flow is strongly suggested by the success of
relativistic 
hydrodynamics to explain particle production at small transverse momenta
\cite{muller,harris}.
Hydrodynamics implies strong position-momentum correlations. At high energies,
energy loss is determined by
momentum transferred perpendicularly to the parton trajectory. Thus, in the
presence of collective flow these momentum exchanges acquire a preferred
direction which will influence gluon radiation off the fast parton (if it is
created decoupled from the medium i.e. not at rest in the local co-moving
frame of the medium), see Fig.
\ref{fig7}.
In \cite{Armesto:2004pt} we have examined the effect of flow on the jet
profile, see Fig. \ref{fig8}.
With the parameters used in this Fig., the induced mean energy loss is 23 GeV,
which is a conservative estimate for the LHC \cite{Salgado:2003rv,andreas}.
Also the magnitude of the directed component, equal in these results
to the isotropic one,
could be considerably larger. The displacement of the calorimetric center of
the jet is small, $\langle \Delta
\eta\rangle=0.04$, and should not spoil the jet definition which is done using
much larger cone sizes. The effect of flow also leads to different jet widths
in different
$\eta-\phi$ directions, a feature which could be explored in experimental
analysis e.g. \cite{Wang:2004kf}, and to a moderate increase in $v_2$. 
A possible influence of flow has to be considered to extract
densities from quenching studies as flow may mimic the effect of a higher
density. It may also help to understand the space-time
picture of the collision.

\section{Remarks}
\label{sec5}

To conclude, let me briefly highlight some points.
First, energy constraints create large uncertainties for
small $p_\perp$, which motivates the computation of
sub-leading
energy corrections to the existing formalism. Second, an implementation of
medium-induced gluon radiation
in a Monte Carlo simulation is needed, see \cite{andreas}. Third, the
opaqueness of the medium at RHIC makes
the determination of densities difficult; flow effects should also be
considered. Fourth, until now the main focus of the phenomenological
analysis is on single particle spectra, but more differential observables like
jet shapes and multiplicities offer valuable information; this demands a
better
understanding of the vacuum (pp and pA). To conclude, at the LHC high-$E_T$ jets
($E_T>50$ GeV) will be
very abundant \cite{Accardi:2003gp}, so jet quenching studies will play a
prominent role in the
heavy ion program \cite{andrea,andreas,rosselet,bolek}.

\bigskip

{\small I thank the organizers, specially K. Safarik,
for their invitation to review this subject
in such a nice meeting, and C.A. Salgado and U.A. Wiedemann for a
pleasant collaboration.}


\bigskip

\begin{thebibliography}{99}

\bibitem{muller} B. Muller: these proceedings.

\bibitem{harris} J.W. Harris: these proceedings.

\bibitem{Baier:2000mf}
R. Baier, D. Schiff, and B.G. Zakharov:
Ann. Rev. Nucl. Part. Sci.  {\bf 50} (2000) 37.

\bibitem{Kovner:2003zj}
A. Kovner and U.A. Wiedemann:
arXiv:hep-ph/0304151.

\bibitem{Gyulassy:2003mc}
M. Gyulassy, I. Vitev, X.N. Wang, and B.W. Zhang:
arXiv:nucl-th/0302077.

\bibitem{Salgado:2003qc}
C.A. Salgado:
Mod. Phys. Lett. A {\bf 19} (2004) 271.

\bibitem{Vitev:2004bh}
I. Vitev:
J. Phys. G {\bf 30} (2004) S791.

\bibitem{Bjorken:1982tu}
J.D. Bjorken:
preprint FERMILAB-PUB-82-059-THY.

\bibitem{Gyulassy:1993hr}
M. Gyulassy and X.N. Wang:
Nucl. Phys. B {\bf 420} (1994) 583.

\bibitem{Baier:1994bd}
R. Baier, Y.L. Dokshitzer, S. Peigne, and D. Schiff:
Phys. Lett. B {\bf 345} (1995) 277.

\bibitem{Wiedemann:2004wp}
U.A. Wiedemann:
J. Phys. G {\bf 30} (2004) S649.

\bibitem{Zakharov:1996fv}
B.G. Zakharov:
JETP Lett.  {\bf 63} (1996) 952.

\bibitem{Wiedemann:2000za}
U.A. Wiedemann:
Nucl. Phys. B {\bf 588} (2000) 303.

\bibitem{Baier:2002tc}
R. Baier:
Nucl. Phys. A {\bf 715} (2003) 209.

\bibitem{Guo:2000nz}
X.f. Guo and X.N. Wang:
Phys. Rev. Lett.  {\bf 85} (2000) 3591.

\bibitem{Baier:2001yt}
R. Baier, Y.L. Dokshitzer, A.H. Mueller, and D. Schiff:
JHEP {\bf 0109} (2001) 033.

\bibitem{Wang:2002ri}
E. Wang and X.N. Wang:
Phys. Rev. Lett.  {\bf 89} (2002) 162301.

\bibitem{Salgado:2003gb}
C.A. Salgado and U.A. Wiedemann:
Phys. Rev. D {\bf 68} (2003) 014008.

\bibitem{andrea} A. Dainese: these proceedings.

\bibitem{Drees:2003zh}
A. Drees, H. Feng and J. Jia:
arXiv:nucl-th/0310044.

\bibitem{Baier:1998yf}
R. Baier, Y.L. Dokshitzer, A.H. Mueller, and D. Schiff:
Phys. Rev. C {\bf 58} (1998) 1706.

\bibitem{Salgado:2002cd}
C.A. Salgado and U.A. Wiedemann:
Phys. Rev. Lett.  {\bf 89} (2002) 092303.

\bibitem{Gyulassy:2000gk}
M. Gyulassy, I. Vitev, and X.N. Wang:
Phys. Rev. Lett.  {\bf 86} (2001) 2537.

\bibitem{Dainese:2004te}
A. Dainese, C. Loizides, and G. Paic:
arXiv:hep-ph/0406201.

\bibitem{Eskola:2004cr}
K.J. Eskola, H. Honkanen, C.A. Salgado, and U.A. Wiedemann:
arXiv:hep-ph/0406319.

\bibitem{Vitev:2004gn}
I. Vitev:
arXiv:nucl-th/0404052.

\bibitem{Adil:2004cn}
A. Adil and M. Gyulassy:
arXiv:nucl-th/0405036.

\bibitem{Wang:2004yv}
X.N. Wang:
arXiv:nucl-th/0405029.

\bibitem{Dokshitzer:1991fc}
Y.L. Dokshitzer, V.A. Khoze, and S.I. Troian:
J. Phys. G {\bf 17} (1991) 1481.

\bibitem{Dokshitzer:2001zm}
Y.L. Dokshitzer and D.E. Kharzeev:
Phys. Lett. B {\bf 519} (2001) 199.

\bibitem{Djordjevic:2003qk}
M. Djordjevic and M. Gyulassy:
Phys. Lett. B {\bf 560} (2003) 37.

\bibitem{Djordjevic:2003zk}
M. Djordjevic and M. Gyulassy:
Nucl. Phys. A {\bf 733} (2004) 265.

\bibitem{Zhang:2003wk}
B.W. Zhang, E. Wang, and X.N. Wang:
Phys. Rev. Lett.  {\bf 93} (2004) 072301.

\bibitem{Armesto:2003jh}
N. Armesto, C.A. Salgado, and U.A. Wiedemann:
Phys. Rev. D {\bf 69} (2004) 114003.

\bibitem{Adcox:2002cg}
K. Adcox {\it et al.}  [PHENIX Collaboration]:
Phys. Rev. Lett.  {\bf 88} (2002) 192303.
%

\bibitem{Adler:2004}
S.S. Adler {\it et al.}  [PHENIX Collaboration]:
arXiv:nucl-ex/0409028.

\bibitem{Batsouli:2002qf}
S. Batsouli, S. Kelly, M. Gyulassy, and J.L. Nagle:
Phys. Lett. B {\bf 557} (2003) 26.

\bibitem{Salgado:2003rv}
C.A. Salgado and U.A. Wiedemann:
Phys. Rev. Lett.  {\bf 93} (2004) 042301.

\bibitem{Abbott:1997fc}
B. Abbott, M. Bhattacharjee, D. Elvira, F. Nang, and H.~Weerts  [D0
Collaboration]:
preprint FERMILAB-PUB-97-242-E.

\bibitem{Armesto:2004pt}
N. Armesto, C.A. Salgado, and U.A. Wiedemann:
arXiv:hep-ph/0405301.

\bibitem{andreas} A. Morsch: these proceedings.

\bibitem{Wang:2004kf}
F. Wang [STAR Collaboration]:
J. Phys. G {\bf 30} (2004) S1299 and talk at Quark Matter 2004
(http://www-rnc.lbl.gov/qm2004/talks/parallel/Friday01/FWang.pdf).

\bibitem{Accardi:2003gp}
A. Accardi {\it et al.}:
arXiv:hep-ph/0310274.

\bibitem{rosselet}  L. Rosselet: these proceedings.

\bibitem{bolek} B. Wyslouch: these proceedings.

\end{thebibliography}
\end{document}